\documentstyle[prd,tighten,aps]{revtex}
\begin{document}
\draft
\twocolumn[\hsize\textwidth\columnwidth\hsize\csname
@twocolumnfalse\endcsname

\title{Gauge fixing and the Hamiltonian for cylindrical spacetimes}
\author{Guillermo A. Mena Marug\'{a}n}
\address{I.M.A.F.F., C.S.I.C., Serrano 121, 28006 Madrid, Spain}
\maketitle

\begin{abstract}

We introduce a complete gauge fixing for cylindrical spacetimes
in vacuo that, in principle, do not contain the axis of
symmetry. By cylindrically symmetric we understand spacetimes
that possess two commuting spacelike Killing vectors, one of
them rotational and the other one translational. The result of
our gauge fixing is a constraint-free model whose phase space
has four field-like degrees of freedom and that depends on three
constant parameters. Two of these constants determine the global
angular momentum and the linear momentum in the axis direction,
while the third parameter is related with the behavior of the
metric around the axis. We derive the explicit expression of the
metric in terms of the physical degrees of freedom, calculate
the reduced equations of motion and obtain the Hamiltonian that
generates the reduced dynamics. We also find upper and lower
bounds for this reduced Hamiltonian that provides the energy per
unit length contained in the system. In addition, we show that
the reduced formalism constructed is well defined and consistent
at least when the linear momentum in the axis direction
vanishes. Furthermore, in that case we prove that there exists
an infinite number of solutions in which all physical fields are
constant both in the surroundings of the axis and at
sufficiently large distances from it. If the global angular
momentum is different from zero, the isometry group of these
solutions is generally not orthogonally transitive. Such
solutions generalize the metric of a spinning cosmic string in
the region where no closed timelike curves are present.

\end{abstract}

\vskip2pc]

\renewcommand{\thesection}{\Roman{section}}
\renewcommand{\theequation}{\arabic{section}.\arabic{equation}}

\section{Introduction}

Vacuum cylindrical spacetimes have received intensive study in
general relativity. The reasons for this interest can be found
in the fact that cylindrical symmetry allows for a rich variety
of physical phenomena while considerably simplifying Einstein's
equations, so that one can obtain non-trivial exact solutions
\cite{MAC}. The first family of time-dependent cylindrical
spacetimes in vacuo seems to have been found by Beck in the
1920's \cite{BE}. This family was rediscovered ten years later
by Einstein and Rosen, in a systematic investigation of all
cylindrically symmetric solutions that describe linearly
polarized radiation \cite{ER}. The most general solution
corresponding to cylindrical gravitational waves in vacuo
(without the condition of linear polarization) was analyzed by
Ehlers and collaborators, and independently by Kompaneets
\cite{KO}. By studying the dynamical equations, Thorne \cite{TH}
succeeded in constructing a covariant vector that satisfies a
conservation law and provides a notion of energy for these
cylindrical spacetimes. This $C$ energy, which is positive and
localizable, is in fact an energy density per unit length along
the axis of symmetry. In the 1970's, Kucha\v{r} discussed the
canonical formalism for Einstein-Rosen waves in the context of
quantum gravity \cite{KU}. More recently, cylindrical
gravitational waves have been considered as a particular case of
spacetimes that possess a translational spacelike Killing field
\cite{AV,AB}. This class of spacetimes can be reduced to three
dimensions using their symmetry. In this way, Ashtekar and
Varadarajan showed that cylindrical waves admit a well-defined
Hamiltonian formalism and that the Hamiltonian that generates
asymptotic time tanslations at spatial infinity is not exactly
the (total) $C$ energy, but a non-polynomial function of it
which, in addition to being positive, turns out to be bounded
also from above \cite{AV,VA}. The same conclusion about the
value and bounds of the Hamiltonian was obtained from a purely
four-dimensional perspective by Romano and Torre \cite{RT} and,
for the particular case of Einstein-Rosen waves, also by
analyzing the asymptotic structure at null infinity of the
three-dimensional Killing reduction of the system \cite{AB}.

The wave solutions in vacuo analyzed in all these works are
obtained by adopting a definition of cylindrical symmetry that
might be considered too restrictive. In addition to the
existence of a translational and a rotational Killing field, it
is assumed that the spacetime manifold contains at least part of
the axis of cylindrical symmetry, namely, the set of fixed
points of the axial Killing field \cite{CSV}. Under such
hypotheses, the geometry must be regular at the axis, and it is
then possible to show that the isometry group generated by the
two Killing vectors is Abelian \cite{CSV} and orthogonally
transitive \cite{CA,BCM,PA}; i.e., the Killing orbits admit
orthogonal surfaces. Obviously, the assumption of regularity
eliminates interesting possibilities that have found
applications in astrophysics and cosmology. This is the case,
e.g., of straight cosmic strings, namely, one-dimensional
topological defects with a linear energy density that introduce
a conical singularity at the axis and, therefore, a deficit
angle in the geometry \cite{CS}. Orthogonal transitivity (a
consequence of the regularity at the axis) precludes as well the
existence of a global rotation \cite{BCM} which is present, for
instance, in spinning string solutions \cite{DJH,SS}. These
solutions have axial singularities produced by string-like
defects that carry a non-zero angular momentum per unit length
in the axis direction and, in principle, may have vanishing
energy density. In the absence of gravitational radiation, the
energy content and angular momentum due to a cosmic string were
analyzed from a three-dimensional viewpoint by Deser, Jackiw,
and 't Hooft \cite{DJH} and also by Henneaux \cite{HE}. On the
other hand, a proposal has been recently made to extend the
concept of energy from Einstein-Rosen waves to orthogonally
transitive spacetimes that contain a non-spinning cosmic string
\cite{HA}. This proposal, nevertheless, is not based on a
Hamiltonian analysis.

The purpose of the present work is to generalize the study of
the Hamiltonian structure and physical degrees of freedom of
vacuum cylindrical solutions to the case in which the axis of
symmetry is not included in the spacetime, so that singularities
can appear there. In more detail, we will assume that there
exists an Abelian two-dimensional group of isometries, generated
by an axial and a translational spacelike Killing field, but we
will not suppose that the axis belongs to the vacuum spacetime
or that the isometry group is orthogonally transitive. Our aim
is to introduce a complete gauge-fixing procedure and analyze
the dynamics of the resulting reduced system. We want to
investigate whether such a reduced dynamics admits a
well-defined Hamiltonian formalism and, if the answer is in the
affirmative, determine whether the existence of upper and lower
bounds for the Hamiltonian still holds when the assumption of
regularity at the axis is dropped.

The rest of the paper is organized as follows. In Sec. II, we
develop a complete gauge fixing for cylindrical spacetimes in
vacuo. For the momentum constraints that correspond to the
Killing fields, the gauge freedom is fixed in Sec. II A. Section
II B introduces a convenient change of metric variables,
suitable for the study of cylindrical spacetimes. The gauge
freedom associated with the remaining momentum constraint is
removed in Sec. II C. Finally, we eliminate the Hamiltonian
constraint in Sec. II D. The reduced system attained in this way
is analyzed in Sec. III. Using the symplectic structure induced
from general relativity, we find in Sec. III A a Hamiltonian
that, at least formally, generates the dynamics of the reduced
model. The explicit expression of the line element in terms of
the four field-like degrees of freedom of the phase space of the
system is presented in Sec. III B. We also include there the
dynamical equations that dictate the evolution of the reduced
model. In Sec. IV we prove that, when the reduced Hamiltonian is
well defined, its range is bounded both from below and above. In
Sec. V we discuss the conditions that ensure that the reduced
Hamiltonian formalism is mathematically consistent and study the
possible divergences of the metric at the axis. Section VI
summarizes the main results of the work and includes some
further comments. Finally, boundary conditions on our physical
fields leading to an acceptable Hamiltonian formalism are
presented in the Appendix.

\section{Gauge Fixing}
\setcounter{equation}{0}

Our starting point is the Hamiltonian formulation of general
relativity. We assume that the spacetime is globally hyperbolic,
so that it admits a 3+1 decomposition in sections of constant
time $t$. In addition, we suppose that there exist two commuting
spacelike Killing vector fields, one of them axial and the other
one translational. Since the isometry group is Abelian (with
non-null orbits), it is possible to choose spatial coordinates
$x^i=\{z,\theta,u\}$ ($i=1,2,3$) such that $\partial_z$ and
$\partial_{\theta}$ are the translational and rotational Killing
fields, respectively, and the spacetime metric is independent of
$z$ and $\theta$ \cite{MAC,BCM}. As a consequence of this
independence, the integral $\int dz \oint d\theta$ appears as a
global factor in the gravitational action and in the symplectic
structure of general relativity. We absorb the numerical value
of $\int dz$ in Newton's constant (by renormalization if $z$ has
infinite length \cite{MM}) and call $G$ the effective
gravitational constant obtained in this manner. In addition, we
normalize the coordinate $\theta$ so that it belongs to the unit
circle $S^1$ (hence, $\oint d\theta=2\pi$) and adopt units such
that $4G=c=1$. As for the spatial coordinate $u$, we choose its
domain of definition equal to the real line. This choice is
always compatible with the fact that $\partial_{\theta}$ is
rotational if one accepts that the axis of symmetry is not
included in our spacetime (think, e.g., of the change $u=\ln{r}$
if $r$ is a radial coordinate).

Our system has the symplectic form
\begin{equation}\label{Sym}
\Omega=\int_{I\!\!\!\,R} du\, {\bf d}\Pi^{ij}\wedge {\bf
d}h_{ij},\end{equation} where ${\bf d}$ and $\wedge$ denote the
exterior derivative and product. In terms of the induced metric
$h_{ij}$ and its extrinsic curvature $K_{ij}$, the canonical
momenta can be written \cite{WA}
\begin{equation}\label{Mom}
\Pi^{ij}=\frac{1}{2}h^{1/2}(h^{ik}h^{jl}-h^{ij}h^{kl})K_{kl},
\end{equation}
with $h$ and $h^{ij}$ being the determinant and the inverse of
the three-metric $h_{ij}$. The non-vanishing Poisson brackets
derived from the above symplectic form are
\begin{equation}\label{Poi}
\{h_{ij}(u),\Pi^{kl}(\bar{u})\}=\delta^{(k}_i\delta^{l)}_j
\delta(u-\bar{u}).\end{equation}
Here, $\delta ^i_j$ and $\delta(u)$ are the Kronecker delta and
the Dirac delta, and the indices in parentheses are symmetrized.
Calling $\tilde{\cal H}$ the densitized Hamiltonian constraint
(i.e., the product of the Hamiltonian constraint by $h^{1/2}$)
and ${\cal H}_i$ the momentum constraints, the time derivative
of any function $F$ on phase space is then given by the formula
\begin{equation}\label{tim} \dot{F}=\partial_t F+\left\{F,\int_{
I\!\!\!\,R} du ( N_{_{_{\!\!\!\!\!\!\sim}}\;}\tilde{\cal H}+
N^i{\cal H}_i)\right\},\end{equation} where the overdot denotes
the time derivative, $\partial_t$ is the partial derivative with
respect to the explicit time dependence, $N^i$ is the shift
vector, and $N_{_{_{\!\!\!\!\!\!\sim}}\;}=h^{-1/2}N$ is the
densitized lapse ($N$ being the lapse function) \cite{WA}.

\subsection{Momentum constraints for the Killing fields}

Let us first fix the gauge freedom associated with the momentum
constraints of the two coordinates $x^a=\{z,\theta\}$ ($a,b=1,2$
from now on). Remembering the independence of the metric on
these coordinates, one can check \cite{BCM,MM} that, in our
system of units,
\begin{equation}\label{Ha} {\cal H}_a=-2(h_{ai}\Pi^{iu})
^{\prime}\;.\end{equation} Here, the prime denotes the
derivative with respect to $u$ and we have introduced the
alternative notation $u$ for the spatial index $i=3$. It is then
possible to remove the corresponding gauge freedom by demanding
that, when restricted to the sections of constant time (and only
then), the action of the isometry group be orthogonally
transitive, namely, that $h_{au}=0$. It is easily seen that
these gauge conditions are second class with the momentum
constraints that we want to eliminate \cite{MM}. In order to
arrive at a consistent gauge fixing, we therefore must only find
values for the shift components $N^a$ such that the conditions
$h_{au}=0$ are stable in the evolution. Using the fact that the
solution to the momentum constraints ${\cal H}_a=0$ is given in
our gauge by $\Pi^{au}=h^{ab}c_b/4$, with $c_a(t)$ being two
real functions of the time coordinate, a trivial calculation
leads to the conclusion that the stability condition
$\dot{h}_{au}=0$ implies
\begin{equation}\label{Na}
N^a(u)=\int_u^{\infty}N_{_{_{\!\!\!\!\!\!\sim}}\;}h_{uu}h^{ab}
c_b.\end{equation} Two additive (time-dependent) integration
constants have been removed from $N^a$ by imposing that these
components of the shift vanish in the limit $u\rightarrow
\infty$ or, equivalently, by a suitable redefinition of the
coordinates $x^a$.

On the other hand, employing Eq. (\ref{tim}), it is possible to
check that the dynamical evolution leaves invariant the
variables $\Pi_a^u$, so that the functions $c_a=4\Pi_a^u$ are in
fact constants. Furthermore, using relation (\ref{Mom}), one can
show that
$c_a=|g|^{-1/2}\tilde{\eta}^{\gamma\mu\nu\sigma}\,^{(1)}X_{\mu}
\,^{(2)}X_{\nu}\,^{(a)}X_{\sigma;\gamma}$ \cite{BCM,GE},
where $g$ is the determinant of the four-metric, the semicolon
stands for covariant derivative, Greek letters denote spacetime
indices, $\tilde{\eta}^{\gamma\mu\nu\sigma}$ is the totally
antisymmetric Levi-Civit\`{a} tensor density, and $^{(a)}X$ are the
two Killing fields, $\partial_z$ ($a=1$) and $\partial_{\theta}$
($a=2$). It therefore follows that the orbits of these Killing
fields admit orthogonal surfaces if and only if $c_a=0$
\cite{MAC,CA}. It is clear that the constants $c_a$ are
intimately related to global properties of the spacetime.
Whenever they are different from zero, the geometry cannot be
regular at the axis and the sections of constant time of the
vacuum spacetime cannot have the topology of $I\!\!\!\,R^3$.
Since we are not assuming orthogonal transitivity, we will not
impose that $c_a$ vanish. Nevertheless, although we will allow
for the presence of non-zero constants $c_a$ in our
calculations, they will be treated as parameters that determine
different sectors of the cylindrical reduction of general
relativity, and not as physical degrees of freedom of the
theory.

\subsection{Change of metric variables}

After the above partial gauge fixing, we will introduce a change
of metric variables that leads to a much more convenient
expression for the line element of our spacetimes with two
commuting Killing fields, namely,
\begin{eqnarray}
&&ds^2=e^{2w+y}\left[-f^2N_{_{_{\!\!\!\!\!\!\sim}}\;}^2dt^2+
(du+N^udt)^2\right]+e^{y}f^2\nonumber\\ \label{met}
&&\times(d\theta+N^{\theta}dt)^2\!+e^{-y}
\left[dz-vd\theta+(N^z\!-\!vN^{\theta})dt\right]^2.
\end{eqnarray}
The new metric variables that replace $h_{uu}$ and the symmetric
two-metric $h_{ab}$ are ${q}\equiv\{f,v,y,w\}$. The restriction
to (inequivalent) positive definite three-metrics $h_{ij}$
requires that $f$ be (e.g.) strictly positive whereas the rest
of metric variables must be real. The momenta $p_{q}$
canonically conjugate to the metric variables $q$ can be easily
found: $p_q=\Pi^{uu}\partial_qh_{uu}+\Pi^{ab}\partial_qh_{ab}$.
Then, the reduced symplectic form of our gauge-fixed model is
$\Omega_{1}=\int du {\bf d}p_q\wedge {\bf d}q$. On the other
hand, the two constraints that remain on the system can be
written
\begin{eqnarray}
&&\tilde{\cal
H}=\frac{(y^{\prime}f)^2}{4}+\frac{(v^{\prime})^2}{4}e^{-2y}+
p_y^2+p_v^2f^2 e^{2y}-fp_wp_f\nonumber\\ \label{coh}
&&\;\;\;\;\;+f(f^{\prime\prime}-f^{\prime}w^{\prime})
+\frac{e^{2w}}{4f^2}\left[(c_{\theta}+c_zv)^2+c_z^2f^2e^{2y}\right]
.\\ \label{cohu} &&{\cal H}_u=-p_w^{\prime}+p_f f^{\prime}+ p_v
v^{\prime}+p_w w^{\prime}+p_y y^{\prime}.\end{eqnarray} Notice
that, when $c_z=c_{\theta}=0$, these formulas reproduce the
results obtained for spacetimes that are regular at the axis
\cite{RT,AG,TO}. Finally, it is possible to check that the
equations of motion obtained from Eq. (\ref{tim}) for the
degrees of freedom of our gauge-fixed model coincide in fact
with those generated in our reduced system by the Hamiltonian
$\int du (N_{_{_{\!\!\!\!\!\!\sim}}\;}\tilde{\cal H}+N^u{\cal
H}_u)$.

\subsection{Radial momentum constraint}

Our next step in the process of gauge fixing consists in
eliminating the gauge freedom associated with the momentum
constraint ${\cal H}_u$. This can be done, e.g., by imposing
that the metric variable $f$ be a fixed, strictly increasing
function of only the coordinate $u$, so that, once the value of
$f$ is known, the coordinate $u$ is totally determined. We note
that, from expression (\ref{met}), $f^2$ is just the determinant
of the metric on Killing orbits. In particular, this metric
degenerates if $f$ vanishes. The set of points where $f=0$,
which are in principle excluded from our spacetime, would then
correspond to the axis of symmetry. Thus, by introducing a
change of coordinates that replaced $u$ with $f$, we could
interpret $f$ as a kind of radial coordinate (recall that $f$ is
positive). We will return to this point in Sec. III B.

Let us hence impose the condition $f=r(u)$, where $r(u)$ is a
fixed function that is strictly positive and increasing, so that
$r(u)>0$ and $r^{\prime}(u)>0$ everywhere. Although the
expression of $r(u)$ is given once and for all, we will not
specify it explicitly; instead, we will treat $r(u)$ as an
abstract fixed function. It is clear that our gauge-fixing
condition does not commute under Poisson brackets with the
constraint ${\cal H}_u=0$. So our gauge fixing will be
acceptable if we can find a value for the shift component $N^u$
such that our choice of gauge is stable, namely, such that
$\dot{f}=0$ on our gauge section. One can see \cite{MM} that
this requirement implies that $N^u=
N_{_{_{\!\!\!\!\!\!\sim}}\;}p_w /(\ln{r})^{\prime}$. On the
other hand, substituting $f=r(u)$ in Eq. (\ref{cohu}), it is
easy to find the expression for $p_f$ that solves the constraint
${\cal H}_u=0$. After removing the degrees of freedom $f$ and
$p_f$ from the system, we arrive at a reduced phase space with
symplectic form
\begin{equation}\label{Om2}\Omega_{2}=\int_{I\!\!\!\,R}
 du \,({\bf d}p_v\wedge {\bf d}v+
{\bf d}p_w\wedge {\bf d}w+{\bf d}p_y\wedge{\bf
d}y).\end{equation} The system has only one constraint, the
densitized Hamiltonian constraint (\ref{coh}) evaluated on our
gauge section, which we will also call $\tilde{\cal H}$. One can
check that the smeared constraint $\int du
N_{_{_{\!\!\!\!\!\!\sim}}\;}\tilde{\cal H}$ generates the
reduced dynamics via the Poisson brackets obtained from
$\Omega_{2}$.

\subsection{Hamiltonian constraint}

In order to complete our gauge fixing, we must remove the gauge
freedom corresponding to the densitized Hamiltonian constraint.
One can use this freedom to impose that the metric induced on
the reference surfaces with coordinates $t$ and $u$ be diagonal.
Since $g_{tu}=g_{uu}N^u$ and, according to our results, $N^u$ is
proportional to $p_w$, it will suffice to demand that $p_w$
vanish. It is not difficult to check that the condition $p_w=0$
is second class with the constraint $\tilde{\cal H}=0$. On the
other hand, the stability of our gauge fixing (i.e.,
$\dot{p}_w=0$) implies that $[\ln{(N_{_{_{\!\!\!\!\!\!\sim}}\;}
rr^{\prime})}]^{\prime}=-e^{2w}G/r^{\prime}$, where
\begin{equation}\label{G}G=\frac{1}{2r^3}\left[
(c_{\theta}+c_zv)^2+c_z^2r^2e^{2y}\right].\end{equation} The
above differential equation provides then a unique value for
$N_{_{_{\!\!\!\!\!\!\sim}}\;}$ under the condition that the
lapse $N$ be asymptotically unity, namely, that
$\lim_{u\rightarrow\infty}N=1$.

In addition, the constraint $\tilde{\cal H}=0$ leads to a
non-linear and inhomogeneous first-order differential equation
for $w$ that, in spite of the apparent complication, can be
solved exactly. The solution for vanishing $p_w$ is
\begin{equation}\label{w}
e^{2w}=\frac{(r^{\prime})^2\exp{\left(\int^u_{u_0}H\right)}}{
(r^{\prime}_0)^2e^{-2w_0}-\int^u_{u_0}d\hat{u}\; r^{\prime}G
\exp{\left(\int^{\hat{u}}_{u_0}H\right)}}.
\end{equation}
Here, $u_0$ is a fixed point, used as the end point in all our
integrations (which are over the dependence on the coordinate
$u$), $w_0(t)=w(u=u_0,t)$, and
\begin{equation}\label{H}
H=\frac{2}{rr^{\prime}}\left[\frac{(ry^{\prime})^2}{4}+
\frac{(v^{\prime})^2}{4}e^{-2y}+p_y^2+p_v^2r^2e^{2y}\right].
\end{equation}
Employing this solution, together with the boundary condition
$\lim_{u\rightarrow\infty}N=1$, it is not difficult to integrate
the differential equation satisfied by the densitized lapse. One
obtains
\begin{equation}\label{N}
N_{_{_{\!\!\!\!\!\!\sim}}\;}=A_{\infty}\frac{r^{\prime}}{r}
e^{-2w}\exp{\left(-\int^{\infty}_uH\right)},\end{equation} with
\begin{equation}\label{Ainf}
A_{\infty}=\frac{e^{w_{\infty}
-y_{\infty}/2}}{r^{\prime}_{\infty}},
\end{equation} $w_{\infty}$, $y_{\infty}$, and
$r^{\prime}_{\infty}$ being the limits of $w$, $y$, and
$r^{\prime}$, respectively, when $u\rightarrow \infty$. It is
worth noting that, in order that $w$ be real, the denominator in
Eq. (\ref{w}) has to be positive for all real values of $u$. We
will discuss this point in detail in Sec. IV.

In the rest of our discussion, we will fix $u_0$ at minus
infinity and call $w_0(t)=\lim_{u\rightarrow-\infty}w(u,t)$.
Furthermore, in order to suppress any explicit time dependence
in the solution for $e^{2w}$ given above, we will suppose that
$w_0$ is actually constant. The assumption $\dot{w}_0=0$
introduces then some consistency conditions in our system.
Employing the fact that $\int du
N_{_{_{\!\!\!\!\!\!\sim}}\;}\tilde{\cal H}$ generates the time
evolution before one performs the gauge fixing discussed in this
subsection, one can see that, on our gauge section,
\begin{equation}\label{wdot}
\frac{d(e^{2w})}{\;dt\;}=2A_{\infty}\exp{\left(-\int
^{\infty}_uH\right)}(p_vv^{\prime}+p_yy^{\prime}).
\end{equation}
Let us now suppose that
$\dot{w}_0(t)=\lim_{u\rightarrow-\infty}\dot{w}(u,t)$. This
commutation of the limit $u\rightarrow-\infty$ and the time
derivative should occur at least for sufficiently smooth
solutions $w(u,t)$ if no material sources are present at minus
infinity that might invalidate the vacuum equation of motion for
$w$. Besides, let us admit that $A_{\infty}$ is finite and that,
in the sector of phase space under consideration, $H$ is
integrable over the real line. Then, the requirement that $w_0$
be constant implies that
\begin{equation}\label{con0}
\lim_{u\rightarrow-\infty}(p_v v^{\prime}+p_y y^{\prime})=0.
\end{equation}
In the following, we assume that this condition is satisfied.
Actually, we will see in Sec. V and in the Appendix that, at
least in certain situations, Eq. (\ref{con0}) is satisfied once
one imposes suitable boundary conditions on the physical degrees
of freedom of the system.

Finally, after the gauge fixing explained here, the symplectic
form induced on phase space is
\begin{equation}\label{Omeg}
\Omega_{3}=\int_{I\!\!\!\,R}
 du \,({\bf d}p_v\wedge {\bf v}+{\bf d}p_y\wedge
{\bf d}y).\end{equation} The system is free of constraints and
its physical degrees of freedom are the canonically conjugate
pairs of fields $(v,p_v)$ and $(y,p_y)$.

\section{Reduced model}
\setcounter{equation}{0}

In this section, we will study the constraint-free system
obtained with our gauge fixing. We first show in Sec. III A that
there exists a reduced Hamiltonian that (at least formally)
generates the time evolution. The explicit form of the spacetime
metric in terms of the physical degrees of freedom is given in
Sec. III B. There, we employ the function $r(u)$ as a radial
coordinate, instead of the spatial coordinate $u$. We also
obtain the dynamical equations for the reduced model and show
that, in order to eliminate a physical ambiguity coming from the
freedom in the choice of origin for $y$, one can fix the value
of $y_{\infty}$ equal to zero.

\subsection{Reduced Hamiltonian}

The equations of motion satisfied by the physical degrees of
freedom of our model can be deduced by recalling that, before
fixing the gauge associated with the Hamiltonian constraint, the
dynamics is generated by the Hamiltonian $\int du
N_{_{_{\!\!\!\!\!\!\sim}}\;}\tilde{\cal H}$ via the Poisson
brackets determined by the symplectic form (\ref{Om2}). Once the
time derivatives of $v$, $y$, and their momenta have been
computed in this way, one can evaluate them at $p_w=0$ and
substitute the values of $w$ and $N_{_{_{\!\!\!\!\!\!\sim}}\;}$
given in Eqs. (\ref{w}) and (\ref{N}). The results are the
dynamical equations that dictate the evolution in our reduced
system. Remarkably, it turns out that such equations can be
directly obtained in the constraint-free system, endowed with
the bracket structure provided by the symplectic form
$\Omega_3$, if one employs as reduced Hamiltonian the following
function on phase space:
\begin{equation}\label{HR1}
H_R=-r^{\prime}_{\infty}e^{-w_{\infty}-y_{\infty}/2}+\;{\rm
const}.\end{equation} Note that, assuming that $y_{\infty}$ is
fixed by the boundary conditions, the only phase-space
dependence of $H_R$ is through $w_{\infty}$, which is obtained
from expression (\ref{w}) in the limit that
$u\rightarrow\infty$. As for the additive constant appearing in
Eq. (\ref{HR1}), it seems natural to fix it so that the
Hamiltonian of flat Minkowski spacetime vanishes. We will come
back to this point later in this section.

The simplest way to show that $H_R$ provides a reduced
Hamiltonian is to check that it leads to the correct equations
of motion. Actually, this can be done after a lengthy but
trivial calculation. It is important to remember that, in the
constraint-free system, all degrees of freedom commute under
Poisson brackets with $w_0$, because we have supposed that this
quantity is a fixed constant. Had we not imposed this
restriction, $w_0$ could have contained a non-trivial
phase-space dependence.

An alternative proof that $H_R$ is the Hamiltonian that
generates the reduced dynamics is the following. Let us call
$\chi_1$ the densitized Hamiltonian constraint and $\chi_2$ the
gauge-fixing condition $p_w=0$, and let $c^{(lm)}(u,\bar{u})$ be
the matrix that satisfies
\begin{eqnarray}
&&\int_{I\!\!\!\,R}d\hat{u}\,
c^{(lm)}(u,\hat{u})\{\chi_{m}(\hat{u}),
\chi_{n}(\bar{u})\}_{P}=\delta_{n}^l
\delta(u-\bar{u})\nonumber \\ \label{cmn}
&&=\int_{I\!\!\!\,R}d\hat{u} \,\{\chi_{n}(u),
\chi_{m}(\hat{u})\}_{P}\;c^{(ml)}(\hat{u},\bar{u}).\end{eqnarray}
Here, the Poisson brackets $\{\;,\;\}_{P}$ are those
corresponding to the symplectic structure (\ref{Om2}), i.e.,
before our choice of gauge for the Hamiltonian constraint. The
indices $l$, $m$, and $n$, on the other hand, can take the
values 1 or 2. Then, after completing the gauge fixing, the
brackets of the physical degrees of freedom
$\{\xi\}\equiv\{v,p_v,y,p_y\}$ with $w$ are \cite{DI}
\begin{eqnarray}\label{Dir}
\{\xi(u),w(\bar{u})\}&=&-\int_{I\!\!\!\,R}d\hat{u}
\int_{I\!\!\!\,R}d\check{u}\,\{\xi(u),\chi_1
(\hat{u})\}_{P}\;c^{(1m)}(\hat{u},\check{u})\nonumber \\
&\times&
\{\chi_m(\check{u}),w(\bar{u})\}_{P},\end{eqnarray}
where we have used $\{\xi,w\}_P=\{\xi,p_w\}_P=0$.

The right-hand side of the above formula must be evaluated on
our gauge section once all Poisson brackets have been computed.
On that section, Eq. (\ref{cmn}) is solved by the matrix
$c^{(11)}=c^{(22)}=0$ and
\begin{equation}\label{c12}
c^{(12)}(u,\bar{u})=-c^{(21)}(\bar{u},u)=\frac{\Theta(\bar{u}-u)
N_{_{_{\!\!\!\!\!\!\sim}}\;}(u)}{N_{_{_{\!\!\!\!\!\!\sim}}\;}
(\bar{u})r(\bar{u})r^{\prime}(\bar{u})},\end{equation} up to the
addition of a function of time to the Heaviside function
$\Theta(\bar{u}-u)$. Such an arbitrary function of time is set
in fact equal to zero by the condition that $w_0$ commute with
the physical degrees of freedom, namely, that the right-hand
side of Eq. (\ref{Dir}) vanish in the limit
$\bar{u}\rightarrow-\infty$. Remember that the Heaviside
function $\Theta(x)$ is unity for $x>0$ and zero otherwise, and
that the densitized lapse is given by expression (\ref{N}).
Substituting the above value for $c^{(12)}$ in Eq. (\ref{Dir}),
taking the limit $\bar{u}\rightarrow\infty$ and recalling that
$N_{_{_{\!\!\!\!\!\!\sim}}\;}rr^{\prime}$ tends to
$e^{-w_{\infty}-y_{\infty}/2}r^{\prime}_{\infty}$, whereas
\begin{equation}\label{dyn}
\dot{\xi}(u)=\int_{I\!\!\!\,R} d\hat{u}\,
N_{_{_{\!\!\!\!\!\!\sim}}\;}(\hat{u})\; \{\xi(u),\tilde{\cal
H}(\hat{u})\}_{P},\end{equation} we arrive at the desired result
$\dot{\xi}=\{\xi,H_R\}$. In doing so, we have also used the fact
that $r^{\prime}_{\infty}$ is a constant given by our gauge
fixing and assumed that $y_{\infty}$ is fixed as a boundary
condition.

Taking into account that, apart from a fixed factor,
$e^{-w_{\infty}}$ generates the reduced dynamics and that $w$,
determined by expression (\ref{w}), is explicitly time
independent, we also see that the quantity $w_{\infty}$ is in
fact a constant of motion: its value remains constant in the
classical evolution, although it may vary from one classical
solution to another. Of course, the same result applies to the
reduced Hamiltonian $H_R$, whose value is thus conserved by the
dynamics of the reduced system.

In arriving at this result, the fact that $w_0$ can be set equal
to a fixed constant plays a fundamental role: otherwise,
$w_{\infty}$ would generally display a non-trivial explicit
dependence on time. Remember that, in the absence of external
sources that could affect the value of $\dot{w}$ at minus
infinity, the assumption that $w_0$ is constant (and $w$ smooth)
amounts to condition (\ref{con0}). In a similar way, assuming
that there exist no external sources at infinity that could
modify the value of $\dot{w}$ when $u\rightarrow\infty$, the
constancy of $w_{\infty}$ turns out to introduce an additional
requirement in our system. Arguments like those presented when
deducing Eq. (\ref{con0}) lead to the conclusion
\begin{equation}\label{coninf}
\lim_{u\rightarrow\infty}(p_v v^{\prime}+p_y
y^{\prime})=0.\end{equation} As happens to be the case with the
analogous condition at minus infinity, it is seen in Sec. V (and
in the Appendix) that, at least in certain situations, this
requirement is satisfied as a consequence of the boundary
conditions imposed on the physical fields.

\subsection{Metric and equations of motion}

Let us now summarize the results obtained so far, but performing
a change of coordinates from $u$ to the positive, strictly
increasing function $r$ introduced in Sec. II C. Notice that,
since the new coordinate $r$ is positive and equal to the
determinant of the metric on Killing orbits, it is possible to
interpret it as a radial coordinate. We will denote the limits
of $r(u)$ when $u$ tends to minus and plus infinity,
respectively, by $r_0$ and $r_{\infty}$. Obviously, we have
$0\leq r_0<r_{\infty}$ and the range of $r$ is
$(r_0,r_{\infty})$. The axis $r=0$ is in principle excluded from
our manifold. The phase space of the reduced model has the
symplectic form
\begin{equation}\label{barOm}
\bar{\Omega}=\int^{r_{\infty}}_{r_0}dr\,({\bf d}P_v\wedge {\bf
d}v+{\bf d}P_y\wedge {\bf d}y),\end{equation} where
$P_v=p_v/r^{\prime}$ and $P_y=p_y/r^{\prime}$. Thus, the system
has four physical degrees of freedom, which are given by the
canonical fields $\{v,P_v,y,P_y\}$.

From our discussion in Sec. II, the spacetime metric can be
expressed in terms of these fields as
\begin{eqnarray}
ds^2&=&e^{2\bar{w}+y}\left[-\bar{N}^2dt^2+dr^2\right]+
e^{y}r^2(d\theta+N^{\theta}dt)^2\nonumber\\ \label{metr}
&+&e^{-y} \left[dz-vd\theta+(N^z\!-\!vN^{\theta})dt\right]^2.
\end{eqnarray}
Here
\begin{eqnarray}\label{barw}
&&e^{2\bar{w}}=\frac{e^{2w}}{(r^{\prime})^2} =\frac{E[r]}{
E[r_0]e^{-2\bar{w}_0}-\int^r_{r_0}d\hat{r}\, G E[\hat{r}]} ,\\
\label{Er}
&&E[r]=\exp{\left(-\int^{r_{\infty}}_r
\bar{H}\right)},\\
 \label{barH}
&&\bar{H}=\frac{2}{r}\left[\frac{(r\partial_r y)^2}{4}\!+\!
\frac{(\partial_rv)^2}{4}e^{-2y}+P_y^2+P_v^2r^2e^{2y}\right],\\
\label{barN}
&&\bar{N}=A_{\infty}e^{-2\bar{w}}E[r],\end{eqnarray} with
$A_{\infty}=e^{\bar{w}_{\infty}-y_{\infty}/2}$,
$e^{-\bar{w}_0}=r^{\prime}_0e^{-w_0}$, and $G$ being defined in
Eq. (\ref{G}). All integrals are over the dependence on the
radial coordinate $r$ and $\partial_r$ denotes the partial
derivative with respect to $r$. In addition, the shift vector is
given by
\begin{eqnarray}N^{z}&=&A_{\infty}\int^{r_{\infty}}_{r}
\frac{d\hat{r}}{\hat{r}^3}\,
(c_{\theta}v+c_zv^2+c_z\hat{r}^2e^{2y})\,E[\hat{r}],
\nonumber
\\ \label{Nthe}N^{\theta}&=&
A_{\infty}\int^{r_{\infty}}_{r}\frac{d\hat{r}}
{\hat{r}^3}\,(c_{\theta}+vc_z)\,E[\hat{r}].\end{eqnarray}

The equations of motion that dictate the dynamics of our reduced
system can be deduced, e.g., from Eq. (\ref{dyn}). One finds
\begin{eqnarray}
&&\dot{v}=2A_{\infty}P_v r e^{2y-2\bar{w}}E[r],\nonumber\\
&&\dot{y}=2A_{\infty}\frac{P_y}{r}e^{-2\bar{w}}E[r],\nonumber\\
&&\dot{P}_v= A_{\infty}\partial_r\!\left(\frac{\partial_r v}{2r}
e^{-2y-2\bar{w}}E[r]\right)\!-A_{\infty}\frac{c_z}{2r^3}
(vc_z+c_{\theta})E[r],\nonumber\\
&&\dot{P}_y=A_{\infty}\partial_r\!\left(\frac{\partial_ry}{2}r
e^{-2\bar{w}}E[r]\right)-\frac{A_{\infty}}{2r}e^{-2\bar{w}}
E[r]\nonumber\\ \label{eqdyn}
&&\hspace*{.9cm}\times\left[c_z^2e^{2y+2\bar{w}}+4P_v^2r^2e^{2y}-
(\partial_r v)^2e^{-2y}\right].
\end{eqnarray}
On the other hand, in the absence of sources that could modify
the time variation of $\bar{w}$ at the end points of the domain
of definition of $r$, the requirement that $\bar{w}$ remain
constant at those points [or, strictly speaking, that
$e^{2\bar{w}}$ does; see Eq. (\ref{wdot})] leads to the
condition
\begin{equation}\label{con}
\lim_{r\rightarrow r_0,\,r_{\infty}}
(P_v \partial_rv+P_y \partial_ry)=0,\end{equation} which is the
analogue of Eqs. (\ref{con0}) and (\ref{coninf}).

From the above equations of motion, it is easy to see that the
Minkowskian solution with boundary condition $\lim_{r\rightarrow
r_{\infty}}y=y_{\infty}$ is obtained by setting $v=P_v=P_y=0$
and $y=y_{\infty}$ when the parameters $c_z$, $c_{\theta}$, and
$\bar{w}_0$ vanish. Using then this flat spacetime as the
solution with respect to which one measures the value of the
reduced Hamiltonian, one gets
$H_R=e^{-y_{\infty}/2}(1-e^{-\bar{w}_{\infty}})$.

Several comments are in order at this stage of our discussion.
First, we remark that Minkowski spacetime is a solution of our
reduced system only if the constants $c_z$, $c_{\theta}$ and
$\bar{w}_{0}$ are equal to zero. These constants are supposed to
be parameters of the system, and not physical degrees of
freedom. So, strictly speaking, Minkowski spacetime cannot be
considered a background solution unless the above parameters
vanish in our model. Nevertheless, we can always decide to
measure the value of the reduced Hamiltonian as referred to its
Minkowskian value. What we are doing in this way is to employ a
universal reference for all of the reduced models that are
obtained with different choices of the parameters $c_z$,
$c_{\theta}$, and $\bar{w}_0$.

Second, we note that, with the boundary condition
$\lim_{r\rightarrow r_{\infty}}y=y_{\infty}$, the asymptotic
norm of the Killing field $\partial_z$ generally differs from
unity. The normalized translational Killing field is given by
$e^{y_{\infty}/2}\partial_z$ instead. This fact must be taken
into account if one wants to define quantities per asymptotic
unit length in the axis direction (such as, e.g., a linear
energy density). In fact, it suffices to redefine the system of
units so that $4Ge^{y_{\infty}/2}=1$, where $G$ is the effective
gravitational constant introduced in Sec. II. One can check
that, for all practical purposes, the only important consequence
of this redefinition is the introduction of a shift in the
origin of $y$ that makes the boundary value $y_{\infty}$ equal
to zero. An equivalent way to see that the value taken by
$y_{\infty}$ is physically irrelevant, so that it can be set to
vanish, is the following. It is not difficult to check from the
expression of the metric that an additive constant in the field
$y$ can always be absorbed by a constant scaling of the
coordinates $z$ and $r$, the fields $v$ and $P_v$, and the
constants $c_z$ and $c_{\theta}$. All four-geometries related by
a shift in $y$ and these scaling transformations are thus
equivalent. In order to eliminate this redundancy, one can then
simply fix the value of $y$ at $r_{\infty}$. For convenience, we
will hence take
\begin{equation}\label{yinf}
y_{\infty}=0,\;\;\;\;\;A_{\infty}=e^{\bar{w}_{\infty}}.\end{equation}
So the Hamiltonian that generates the dynamics of the reduced
model can now be written in the form
\begin{equation}\label{Hredu}
H_R=1-e^{-\bar{w}_{\infty}}.\end{equation}

Finally, we notice that, when the constants $c_z$, $c_{\theta}$,
and $\bar{w}_0$ vanish, the formulas given above for the
spacetime metric and reduced Hamiltonian reproduce the results
obtained in the literature for cylindrical waves that are
regular at the axis \cite{RT,AG,AP}.

\section{Energy bounds}
\setcounter{equation}{0}

Let us now show that the linear energy density contained in our
system, which is given by the value of the reduced Hamiltonian,
is bounded both from above and below, like in the case with
regular axis \cite{AV}. In doing this, we will only assume that
the Hamiltonian formalism is well defined in our reduced model.

In the real phase space of our model, the spacetime metric
(\ref{metr}) describes the 3+1 decomposition of a Lorentzian
spacetime with time coordinate $t$ if and only if $\bar{w}$
[given by Eq. (\ref{barw})] is real. In particular, the reduced
Hamiltonian will not generate time evolution unless
$\bar{w}_{\infty}$ is real. So $e^{-\bar{w}_{\infty}}$ must be
strictly positive. As a consequence, we conclude that the
Hamiltonian $H_R$ is bounded from above by unity, $H_R<1$. Here,
we have ruled out the possibility $e^{-\bar{w}_{\infty}}=0$ by
requiring that metric (\ref{metr}) be well defined.

In order to find a lower bound for the Hamiltonian, let us first
note that the quantities $G$ and $\bar{H}$ that enter the
expression of $\bar{w}$ are positive functions on the real phase
space, as can be easily seen from their definitions, taking into
account that the radial coordinate $r$ is positive. It is then
straightforward to check that $e^{2\bar{w}}$ is a strictly
increasing function of $r$, provided that $\bar{w}$ is actually
real. Obviously, this implies that $\bar{w}_{\infty}\geq
\bar{w}_0$. Therefore, for each fixed value of the constant
parameter $\bar{w}_0$, the reduced Hamiltonian is also bounded
from below: $H_R\geq (1-e^{-\bar{w}_0})$.

When the axis of symmetry is regular, so that $\bar{w}_0$
vanishes, we recover the result $1>H_R\geq 0$ \cite{AV}.
Furthermore, assuming as a boundary condition (see Sec. V and
the Appendix for a detailed discussion) that $v$ is much smaller
than the unit function in the limit that $r$ tends to
$r_{\infty}$ and remembering that the shift vector vanishes
asymptotically, one can see that, in this asymptotic region,
metric (\ref{metr}) describes a conical geometry with deficit
angle equal to $2\pi H_R$ (and possibly non-zero angular
momentum). Imposing that the deficit angle be positive amounts
thus to demanding positivity of the energy. From our discussion
above, this positivity could be ensured, e.g., by restricting
the constant parameter $\bar{w}_0$ to be non-negative, because
then $e^{\bar{w}_{\infty}}\geq e^{\bar{w}_0}\geq 1$.

Finally, let us notice that, since $e^{2\bar{w}}$ is strictly
increasing with $r$ (as far as it is positive) and can be seen
to change sign at most once in the positive real axis, the
requirement that $\bar{w}$ be real in the domain of definition
of $r$ is satisfied if and only if $e^{\bar{w}_{\infty}}>0$.
This last condition is stable under dynamical evolution, because
$e^{\bar{w}_{\infty}}$ is a constant of motion. On the other
hand, using formula (\ref{barw}) and recalling that
$e^{\bar{w}_{\infty}}$ must be finite, we can rewrite the
considered condition as
\begin{equation}\label{conw}
E[r_0]e^{-2\bar{w}_0}> \int^{r_{\infty}}_{r_0} dr\,
G\,E[r].\end{equation} The above inequality can be understood as
a restriction on the acceptable values of the phase space
variables for each fixed value of $\bar{w}_0$. In the case that
$c_{z}$ and $c_{\theta}$ vanish, the inequality is trivially
satisfied, because $G$ is then equal to zero.

In conclusion, condition (\ref{conw}) implies the reality of the
metric function $\bar{w}$ everywhere in spacetime and guarantees
that the range of the reduced Hamiltonian is contained in
$[1-e^{-\bar{w}_0},1)$, regardless of the specific values taken
by the parameters $c_{z}$ and $c_{\theta}$ of the model.

\section{Consistency of the formalism and spinning solutions}
\setcounter{equation}{0}

To some extent, the analysis presented in the previous sections
is only formal. The emphasis has been put on removing all the
gauge freedom and finding the expressions of the reduced metric
and Hamiltonian, rather than on proving that such expressions
are well defined. Our aim in this section is to show that, at
least in certain situations, the reduced formalism that we have
discussed is in fact fully consistent.

Let us summarize the conditions that are necessary for the
consistency of the model. First, the metric expression
(\ref{metr}) must be meaningful everywhere. This implies that
the integral
\begin{equation}\label{Ir}
I[r]=\int_{r_0}^rd\hat{r}\,G\,E[\hat{r}]\end{equation} and those
that appear in $E[r]$, $N^z$, and $N^{\theta}$ must converge for
all $r\in (r_0,r_{\infty})$. In addition, one must demand that
$I[r_{\infty}]$ be finite, so that $e^{2\bar{w}_{\infty}}$ is
well defined. Remember also that this constant of motion has to
be positive if the induced metric is positive definite and the
reduced Hamiltonian real. This last condition is equivalent to
requirement (\ref{conw}), which can be interpreted as a
dynamically stable restriction on phase space and implies, in
particular, that $E[r_0]>0$. On the other hand, in order for the
Hamiltonian formalism to be well defined, the reduced
Hamiltonian must not only be real and finite, but also
differentiable on phase space, i.e., with respect to variations
of the fields $v$, $P_v$, $y$, and $P_y$. Other consistency
conditions that must be satisfied are those given in Eq.
(\ref{con}) (which guarantee that $\bar{w}$ is constant at $r_0$
and $r_{\infty}$) and that $y_{\infty}$ can be kept equal to
zero. Finally, note that these requirements must hold at all
instants of time; i.e., the imposed conditions must be stable.

\subsection{General case}

We will first consider the possibility
$0<r_0<r_{\infty}<\infty$. Assuming that all fields are
sufficiently smooth, the integrals that determine the metric
components are then convergent, because they contain no
singularities and the interval of integration is bounded. As for
the differentiability of the reduced Hamiltonian, a detailed
calculation shows that the variation of $H_R$ includes two types
of contributions. The first type consists of integrals over
$r\in(r_0,r_{\infty})$ that converge because of the reasons
explained above. The second type are surface terms that arise in
the integration by parts of the variations of $\partial_r v$ and
$\partial_r y$. These terms must vanish if the Hamiltonian is
differentiable. Notice that the derivatives $\partial_rv$ and
$\partial_ry$ appear in $H_R$ only via the phase-space
dependence of $\bar{H}$, Eq. (\ref{barH}), and that there is no
functional dependence on $\partial_r P_v$ and $\partial_r P_y$.
At least for variations of $v$ and $y$ that are proportional to
the Hamiltonian variations $\dot{v}$ and $\dot{y}$ at $r_0$ and
$r_{\infty}$, a careful analysis proves that the considered
surface terms vanish as a consequence of Eqs. (\ref{con}). In
this sense, in order to guarantee that the reduced formalism is
rigorously defined, one would only need to impose inequality
(\ref{conw}) at a certain instant of time and conditions
(\ref{con}) and $y_{\infty}=0$ for all values of $t$, so that
these last requirements are stable.

However, there is no obvious way in which condition
$y_{\infty}=0$ and Eqs. (\ref{con}) can be satisfied and
preserved in the evolution. Of course, one could assume that
there exist external sources acting on the boundaries of the
spacetime that invalidate the applicability of the reduced
equations of motion (\ref{eqdyn}) at $r=r_0$ and $r=r_{\infty}$.
The possibility of regaining consistency in this way will not be
explored here. There still exists another situation in which our
consistency conditions can be satisfied, namely, in solutions
whose fields $v$ and $y$ are constant (both with respect to $t$
and $r$) outside a certain region of the form $r\in (r_1,r_2)$,
where $r_0<r_1\leq r_2<r_{\infty}$. Of course, we require that
the value of $y_{\infty}$ vanish. Using the dynamical equations
(\ref{eqdyn}), one can easily construct solutions of this type
provided that the constant $c_z$ is equal to zero: it suffices
to assume that the momenta vanish for $r$ in
$(r_0,r_1]\cup[r_2,r_{\infty})$. Note, nevertheless, that such
solutions can always be extended to the whole region
$r\in(0,\infty)$ by keeping the fields $\{v,P_v,y,P_y\}$
constant outside $(r_1,r_2)$.

We are thus naturally led to consider the case $r_0=0$ and
$r_{\infty}=\infty$, either because otherwise we cannot ensure
the consistency of the formalism or because the only interesting
solutions when $r$ has a bounded domain of definition can be
trivially extended to the semiaxis $(0,\infty)$. We will first
analyze models with non-vanishing parameter $c_z$. The
discussion for $c_z=0$ will be presented in the next subsection.

When $c_z\neq 0$, the conditions that $I[\infty]$ be finite and
$E[0]$ positive imply
\begin{equation} \label{Iinf}\infty>\frac{I[\infty]}{E[0]}\geq
\int^{\infty}_0\frac{dr}{2r^3}c_z^2\left[\left(v+\frac{c_{\theta}}
{c_z}\right)^2+r^2e^{2y}\right].\end{equation} In the last
inequality, we have used that $E[r]$ increases with $r$, since
$\bar{H}$ is a positive function on phase space. The convergence
of the last integral would require that
\begin{equation} \label{czli}
\lim_{r\rightarrow 0,\infty} e^y=\lim_{r\rightarrow 0,\infty}
\frac {c_zv+c_{\theta}}{r}=0,\end{equation}
where the limits are taken both at zero and at infinity.
Clearly, the first of these conditions cannot be satisfied
(remember, in particular, that we have assumed $y_{\infty}=0$).
This means that there exist divergences in the metric functions
when $c_z$ does not vanish. More explicitly, the denominator in
$e^{2\bar{w}}$ diverges. Furthermore, it is not difficult to
realize that the divergent terms in $I[r]/E[r_0]$ when
$r_0\rightarrow 0$ cannot actually be absorbed by a kind of
renormalization of the constant parameter $e^{-2\bar{w}_0}$ that
appears in Eq. (\ref{barw}), because those terms depend on the
behavior of the fields $v$ and $y$ around the axis $r=0$ and
vary, in general, from one solution to another.

\subsection{Case $c_z=0$}

Let us now consider the only remaining possibility, i.e., the
case in which the domain of definition of $r$ is the whole
semiaxis $(0,\infty)$ and the constant parameter $c_z$ vanishes.
Using Eq. (\ref{Ha}) and a line of reasoning similar to that
discussed in Refs. \cite{AV,HE}, it is easy to show that
$2\Pi_a^u=c_a/2$ (where we have used the notation of Sec. II A)
is precisely the value of the surface term that must be added to
the smeared momentum constraint $\int du N^iH_i$ in order to
make it differentiable on phase space when the shift vector
$N^i$ equals $\delta^i_a$ ($a=1$ or 2) in the asymptotic region
$u\gg 1$ and vanishes for $u\ll -1$. Therefore, with our choice
of units, $c_{\theta}/2$ and $c_z/2$ are the values per unit
length of the angular momentum and the linear momentum in the
$z$ direction, respectively. The solutions that we are going to
study can thus be regarded as spacetimes with a possibly
singular axis of symmetry, a vanishing linear momentum in the
direction of this axis and, in general, a non-zero angular
momentum.

If we now analyze the behavior of $I[r]$ when $r_0$ approaches
the origin, we easily see that this integral still diverges,
like when $c_z$ differed from zero. However, the leading term in
$I[r]/E[r_0]$ is now the same for all solutions in our model and
can be absorbed in the denominator of $e^{2\bar{w}}$ by a
renormalization of the constant $e^{-2\bar{w}_0}$. Moreover, if
one assumes boundary conditions such that
\begin{equation}\label{Hbc}\lim_{r\rightarrow 0}\,
\frac{\bar{H}}{r^{1+\epsilon}}=0\end{equation}
for a certain number $\epsilon>0$, one can check that the only
divergent term in $I[r]/E[r_0]$ when $r_0$ tends to zero has the
form $c_{\theta}^2/(4r_0^2)$ and is thus universal. So this term
can be removed by redefining $e^{-2\bar{w}_0}=D
+c_{\theta}^2/(4r_0^2)$ and taking the limit $r_0\rightarrow 0$,
where $D$ is a constant parameter. Expression (\ref{barw}) can
then be rewritten
\begin{eqnarray} \label{nbarw}
e^{2\bar{w}}&=&\frac{4\bar{E}[r]r^2}{c_{\theta}^2+4Dr^2-2
c_{\theta}^2r^2\int_0^r d\hat{r}\,\hat{r}^{-3}\,
(\bar{E}[\hat{r}]-1)},\\
\bar{E}[r]&=&\frac{E[r]}{E[0]}=\exp{\left(\int_0^r\bar{H}\right)}.
\end{eqnarray}

Several comments are in order at this point. First, notice that
$\bar{E}[r]$ is a strictly increasing function of $r$ that is
always equal or greater than unity, because $\bar{H}$ is
positive. In addition, assuming that the fields $v$, $P_v$, $y$,
and $P_v$ are sufficiently smooth in the region $0<r<\infty$,
condition (\ref{Hbc}) guarantees that the integrals appearing in
$\bar{E}[r]$ and in the denominator of $e^{2\bar{w}}$ are well
defined and convergent for $r\in [0,\infty)$. On the other hand,
the condition that $E[0]$ be positive amounts to requiring that
$\bar{E}[\infty]$ be finite. This is ensured, e.g., when there
exists a strictly positive number $\delta>0$ such that
\begin{equation}\label{Hinf}
\lim_{r\rightarrow\infty} r^{1+\delta}\bar{H}=0.\end{equation}
Imposing this asymptotic behavior, it is straightforward to
check that $e^{2\bar{w}_{\infty}}$ is well defined and positive
if
\begin{equation}\label{nconw}
D>c_{\theta}^2\int_0^{\infty}\frac{dr}{2r^3}(\bar{E}[r]-1).
\end{equation}
This requirement replaces Eq. (\ref{conw}), owing to the
redefinition of $\bar{w}_0$. In particular, since
$\bar{E}[r]\geq 1$, it is necessary that $D$ be positive. We
will thus restrict our discussion to the case $D>0$ from now on.
Note also that, since $e^{2\bar{w}_{\infty}}$ is a constant of
motion, inequality (\ref{nconw}) is preserved in the evolution.

Concerning the shift vector, the integrals in Eq. (\ref{Nthe})
are meaningful for all $r\in (0,\infty)$ if $v$ has a finite
limit at infinity, which we will assume to vanish (so that the
asymptotic metric describes a conical geometry with possibly a
non-zero angular momentum). Hence, all metric functions are well
defined everywhere in spacetime. Furthermore, from Eqs.
(\ref{Hbc}) and (\ref{barH}), it follows that $\partial_r v$ is
much smaller than $r^{1+\epsilon/2}$ close to the origin. Then,
supposing that $v$ is smooth enough in that region, it is not
difficult to check that $N^z-vN^{\theta}$ has a finite limit
when $r\rightarrow 0$. On the other hand, we can rewrite
$N^{\theta}$ as
\begin{equation}\label{nNthe}
N^{\theta}=e^{\bar{w}_{\infty}}E[0]c_{\theta}\left\{\frac{1}{2r^2}+
\int^{\infty}_r\frac{d\hat{r}}{\hat{r}^3}(\bar{E}[\hat{r}]-1)\right
\},\end{equation}
so that this shift component diverges at the axis $r=0$. From
expressions (\ref{metr}) and (\ref{barN}) we then see that the
only potentially divergent terms in the four-metric when
$r\rightarrow 0$ are included in the diagonal $t$ component, and
are given by
\begin{equation}
-e^y\left\{e^{2\bar{w}_{\infty}}e^{-2\bar{w}}(E[r])^2-r^2(N^
{\theta})^2\right\}.\end{equation} Nevertheless, taking into
account that condition (\ref{Hbc}) guarantees that $E[r]-E[0]$
is much smaller than $r^{2+\epsilon}$ for $r\rightarrow 0$, one
can check that the value obtained from Eqs. (\ref{nbarw}) and
(\ref{nNthe}) for these presumably divergent terms is in fact
finite when $r$ vanishes. Thus, in our coordinate system, the
metric components are well defined even in the limit in which
one reaches the axis of symmetry.

So far we have already proved that, when $c_z=0$ and
$c_{\theta}\neq 0$, there is no problem with the expressions of
the metric and the reduced Hamiltonian, assuming that conditions
(\ref{Hbc}), (\ref{Hinf}), and (\ref{nconw}) are satisfied and
$v$ and $y$ vanish at infinity. In addition, the consistency of
our formalism implies Eqs. (\ref{con}) and the differentiability
of the reduced Hamiltonian. Finally, all these conditions must
be stable. In the Appendix, we present boundary conditions on
the fields $\{v,P_v,y,P_y\}$ at the axis and at infinity that
ensure that all these requirements are satisfied. Nonetheless,
in order to demonstrate the relevance of the reduced model, it
actually suffices to show that the set of acceptable solutions
is infinite dimensional.

In fact, this last statement can be easily proved. As we have
already commented, when $c_z=0$, the dynamical equations
(\ref{eqdyn}) admit sufficiently smooth (even $C^{\infty}$)
solutions in which all fields are constant outside a bounded
interval for $r$ of the form $(r_1,r_2)$, where $r_1$ and $r_2$
satisfy $0<r_1\leq r_2<\infty$ but are otherwise arbitrary.
Besides, outside $(r_1,r_2)$ the momenta $P_v$ and $P_y$ vanish.
The fields $v$ and $y$ are set equal to zero in the interval
$[r_2,\infty)$, so that the condition that these fields vanish
asymptotically is satisfied. To avoid topological complications
on the sections of constant time in the neighborhood of the
axis, we will also assume that $v$ vanishes in the region $r\in
(0,r_1]$. Finally, $y$ will take a constant, finite value $y_0$
in that region. For this infinite family of solutions it is
straightforward to see that all the conditions necessary for
consistency are satisfied, including stability, except maybe the
differentiability of the reduced Hamiltonian and Eq.
(\ref{nconw}). The differentiability of the Hamiltonian can be
checked following a line of reasoning similar to that explained
in the beginning of Sec. V A. The surface terms that appear in
the variation of the Hamiltonian vanish because so do
$\{\partial_r v,P_v,\partial_r y, P_y\}$ at the axis and at
infinity. The remaining terms are integrals that converge
because they get no contribution outside the bounded region
$(r_1,r_2)$, where all integrands are sufficiently smooth.
Hence, the variation is well defined and the reduced Hamiltonian
differentiable. We are only left with condition (\ref{nconw}),
which must be regarded as a restriction on phase space which is
satisfied always in the evolution if so is at a single instant
of time.

It is easy to see that, for any strictly positive constant $D$,
there exists an infinite dimensional set of initial values for
our fields $\{v,P_v,y,P_y\}$ such that inequality (\ref{nconw})
holds. Actually, the minimum of the right-hand side in that
inequality is just zero and is reached when $\bar{H}$ vanishes.
Given expression (\ref{barH}) and that
$y_{\infty}=v_{\infty}=0$, this occurs only for the solution
with vanishing fields. This flat solution can be taken as a
background for our model with fixed parameters $c_{\theta}$ and
$D>0$. The background metric adopts the expression
\begin{equation}\label{scs}
ds^2\!=\!-dt^2\!+\!\frac{c_{\theta}}{\sqrt{D}}dtd\theta+r^2
d\theta^2+dz^2\!+\!\frac
{4r^2dr^2}{c_{\theta}^2+4Dr^2},\end{equation} which is precisely
the line element originated by a spinning cosmic string,
restricted to the region where causality is preserved and there
exist no closed timelike curves (CTC's) \cite{DJH,SS}. A more
familiar form for this metric, which can be continued to the
region $-c_{\theta}^2/(4D)<r^2\leq 0$ at the cost of introducing
CTC's, is obtained with the change of coordinate
$D\rho^2=r^2+c_{\theta}^2/(4D)$. Condition (\ref{nconw}) is
clearly satisfied by our background solution, and one can check
that it is satisfied as well at least in a certain region of
phase space around the origin $v=P_v=y=P_v=0$. Therefore, the
set of admissible spinning solutions is infinite dimensional.

Finally, let us note that, when $c_z=0$, the lower bound
obtained for the reduced Hamiltonian in Sec. IV can be improved.
From Eq. (\ref{nbarw}) and the fact that $\bar{E}[r]\geq 1$, one
gets $e^{-2\bar{w}_{\infty}}\leq D$. We then conclude that the
value of the reduced Hamiltonian, which provides the energy per
unit length in the axis direction, is always contained in the
interval $[1-\sqrt{D},1)$. On the other hand, as we have
commented on, metric (\ref{metr}) describes in the asymptotic
region a conical geometry with angular momentum proportional to
$c_{\theta}$ and deficit angle equal to $2\pi H_R$. Hence,
positivity of the energy and the deficit angle can be ensured,
e.g., by simply restricting the parameter $D$ so that $1\geq
D>0$.

\section{Conclusions and further comments}
\setcounter{equation}{0}

We have proposed a gauge-fixing procedure that removes all the
non-physical degrees of freedom in vacuum cylindrical
spacetimes. Our definition of cylindrical symmetry is less
restrictive than that usually employed in the literature, in the
sense that we have assumed the existence of two commuting
spacelike Killing fields, one of them rotational and the other
one translational, but we have not imposed the condition that
the spacetime contain the axis of rotational symmetry, namely,
the set of points where the metric on Killing orbits
degenerates. This relaxation of the conditions for cylindrical
symmetry has allowed us to include in our discussion spacetimes
whose Killing orbits are not surface orthogonal, so that the
line element cannot be written, in general, in block-diagonal
form using two-metrics. The price to be paid for this
generalization is that now the axis of symmetry, which is
located in principle outside the manifold, may actually be
singular and contain linear sources.

Our gauge fixing leads to a reduced midisuperspace model that is
totally free of constraints and depends on three constant
parameters. Two of these parameters, namely, $c_z$ and
$c_{\theta}$, determine, respectively, the constant values of
the linear momentum in the axis direction and the angular
momentum of the system. The third parameter, $\bar{w}_0$, is the
fixed limit when one approaches the axis ($r\rightarrow r_0$) of
$\bar{w}$, the metric function that appears in the radial
component of the line element (\ref{metr}). The phase space of
the reduced model is infinite dimensional and can be described
using the set of canonically conjugate fields $\{v,P_v,y,P_y\}$.
We have obtained the general expression of the four-metric in
terms of these physical degrees of freedom and found the
equations of motion that govern the evolution of these
independent fields. Moreover, we have proved that the dynamics
of the model is in fact generated by a reduced Hamiltonian,
given by $1-e^{-\bar{w}_{\infty}}$. Here, $\bar{w}_{\infty}$ is
the limit of the metric function $\bar{w}$ at large distances
from the axis ($r\rightarrow r_{\infty}$). The value of this
Hamiltonian is a constant of motion that provides the amount of
energy that is present in the system per unit length in the axis
direction. The origin of energy has been chosen to vanish for
flat, Minkowski spacetime.

One might wonder whether the expression of the reduced
Hamiltonian could also have been obtained from the
Hilbert-Einstein action supplemented with boundary terms via a
reduction process. Actually, the answer turns out to be in the
affirmative, but only if the surface terms are suitably chosen.
One can start with the Hamiltonian form of the gravitational
action corrected with the standard surface terms that appear
when the manifold has a timelike boundary \cite{HH}. In our
case, this boundary consists of two disconnected parts: an
internal boundary at $r=r_0$ and an external one at
$r=r_{\infty}$ (if necessary, one can take the limits
$r_0\rightarrow 0$ and $r_{\infty}\rightarrow
\infty$ after completing all calculations). It is then possible
to show that, if one only includes the surface terms that
correspond to the external boundary, the reduction explained in
Sec. II leads to
\begin{equation}
S_R=\int dt
\left[e^{-\bar{w}_{\infty}}-1+\int_{r_0}^{r_{\infty}}dr(P_v
\dot{v}+P_y\dot{y})\right],\end{equation}
which is in fact the action expected for the reduced system. The
integral over $r$ determines the sympletic structure, whereas
the other factors provide the reduced Hamiltonian. Note that we
have normalized the action so that it vanishes for Minkowski
spacetime. In the case that the axis of symmetry is regular,
which happens only if $c_z$, $c_{\theta}$, and $\bar{w}_0$
vanish, the surface terms at $r_0=0$ that have been obviated are
in fact spurious, because the internal boundary does not exist.
But in the general, singular case, we really need to exclude
those surface corrections in order to arrive at the correct
reduced action. Since the action obtained after reduction
depends on the choice of boundary terms, it is clear that the
symmetric criticality principle does not generally hold in the
system \cite{TO}.

We have also analyzed in detail the conditions that guarantee
that the reduced formalism is consistent. In particular, we have
discussed under what circumstances the metric expressions are
always well defined and the reduced Hamiltonian is real, finite
and differentiable on phase space. In addition, we have checked
whether one can safely impose that, at all instants of time, the
fields $y$ and $v$ vanish asymptotically and Eqs. (\ref{con})
hold. These equations are necessary to ensure that the parameter
$\bar{w}_0$ and the value of the reduced Hamiltonian are
constant. We have proved that, when the radial coordinate $r$ is
defined over the whole semiaxis $(0,\infty)$ and there is no
linear momentum in the direction of the symmetry axis, all the
consistency requirements are satisfied provided that the fields
$\{v,P_v,y,P_y\}$ are subject to appropriate boundary
conditions. We have then particularized our study to models with
$r\in(0,\infty)$ and $c_z=0$ but, in general, with non-vanishing
angular momentum, $c_{\theta}\neq 0$.

For such models, the only apparent problem is a divergence in
the denominator of $e^{2\bar{w}}$ in Eq. (\ref{barw}) when
$r_0\rightarrow 0$. We have shown, however, that this divergence
can be absorbed by a redefinition of the constant
$e^{-2\bar{w}_0}$. We have called $D$ the renormalized constant,
which must be strictly positive. After this redefinition of
parameters, the metric functions are not only well defined
everywhere in spacetime; in addition, with our choice of
coordinates, all metric components turn out to have a finite
limit when the axis of symmetry is approached. Assuming boundary
conditions like those given in the Appendix, the reduced
formalism is fully consistent. Besides, the reduced Hamiltonian,
which determines the linear energy density contained in the
system, is then bounded both from above and below, like in the
case with a regular axis of symmetry \cite{AV}. More explicitly,
in each of the models with constant parameters $c_{\theta}$ and
$D$ (with $c_z=0$), the range of the reduced Hamiltonian is
included in the semi-open interval $[1-\sqrt{D},1)$.
Furthermore, if the deficit angle in the asymptotic region $r\gg
1$ is positive, so must be the energy density per unit length
along the axis.

In the models with $r\in(0,\infty)$ and $c_z=0$ but, possibly, a
non-vanishing angular momentum, a particularly interesting set
of solutions is provided by the following family. We consider a
bounded interval $(r_1,r_2)$, with $0<r_1\leq r_2<\infty$, and
fields that, at a certain instant of time $t=t_0$, satisfy the
conditions that (1) $v$, $P_v$, and $P_v$ vanish outside the
region $r\in(r_1,r_2)$, (2) $y$ be constant for $r$ in $(0,r_1]$
and vanish in $[r_2,\infty)$, (3) Eq. (\ref{nconw}) be
satisfied, and (4) the fields be sufficiently smooth as
functions of $r$ (let us say $C^{\infty}$). These requirements
on the fields are in fact stable in the evolution. One can then
check that all conditions necessary for the consistency of the
reduced formalism are satisfied on these solutions.

Note that we can regard the values of $\{v,P_v,y,P_y\}$ at
$t=t_0$ just as initial data that can be evolved by integrating
(e.g., by numerical methods) the dynamical equations
(\ref{eqdyn}). As we have commented, the result of this
integration is a solution satisfying conditions (1)-(4) at all
instants of time. In this way, one can actually obtain an
infinite number of solutions whose isometry group is not
orthogonally transitive (unless $c_{\theta}=0$).

In addition, it is possible to show \cite{MN} that, at short
distances from the axis, $r\ll 1$, all of these solutions
approach the line element corresponding to a spinning cosmic
string with angular momentum per unit length equal to
$c_{\theta}/2$ and deficit angle given by $2\pi(1-\sqrt{D})$.
Indeed, the metric of this string in the region where no CTC's
are present can be obtained by simply setting the fields
$\{v,P_v,y,P_y\}$ equal to zero [see Eq. (\ref{scs})]. As a
consequence, one can view the metric of the spinning cosmic
string as a flat background for the considered family of
solutions in the model with constant values of the parameters
$c_{\theta}$ and $D$.

\acknowledgments

The author is greatly thankful to N. Manojlovi\'c for valuable
discussions and comments. He is also grateful to J. M. M.
Senovilla for helpful comments. This work was supported by funds
provided by DGESIC under the Research Projects No. PB97-1218 and
No. HP1988-0040.

\section*{Appendix}

In this appendix, we present suitable boundary conditions for
the fields $\{v,P_v,y,P_y\}$ of the reduced model with vanishing
parameter $c_z$ and $r\in(0,\infty)$. The proof that these
conditions are stable under dynamical evolution and that they
ensure the consistency of the reduced Hamiltonian formalism will
be given elsewhere \cite{MN}.

The conditions at infinity, $r\rightarrow\infty$, are that the
field $v$ vanish and that
\[
P_v=O(r^{-1}),\;\;\;\;\;y=O(1),\;\;\;\;\;P_y=O(1).
\]
The notation $f=O(g)$ means that there exists a strictly
positive number $\varepsilon>0$ such that the function $f$ is
much smaller than $r^{-\varepsilon}g$ in the asymptotic region,
i.e., $\lim_{r\rightarrow\infty}r^{\varepsilon}f/g=0$. Note that
the above conditions imply, in particular, that $y_{\infty}=0$.

On the other hand, if the constant parameter $c_{\theta}$
vanishes, we can impose the following conditions in the vicinity
of the axis $r=0$:
\[
v=o(r^2),\;\;\;P_v=o(r),\;\;\;y=y_0(t)+o(r^2),\;\;\;P_y=o(r).
\]
Here, $y_0(t)$ is a time-dependent function, we have assumed
that $v$ vanishes at the axis, and the notation $f=o(g)$ is
employed for functions whose quotient $f/g$ has a finite limit
when $r\rightarrow 0$. Finally, in the case with non-vanishing
global angular momentum, $c_{\theta}\neq 0$, an appropriate
behavior around the axis $r=0$ is
\[
v=o(r^4),\;\;\;P_v=o(r^5),\;\;\;y=y_0(t)+o(r^6),\;\;\;P_y=o(r^3).
\]

\end{document}